# Universal superconductivity phase diagram for pressurized tetradymite topological insulators

Shu Cai[1,4], S. K. Kushwaha[2,8], Jing Guo[1], Vladimir A. Sidorov[3], Congcong Le[1,5], Yazhou Zhou[1], Honghong Wang[1,4], Gongchang Lin[1,4], Xiaodong Li[6], Yanchuan Li[6], Ke Yang[7], Aiguo Li[7], Qi Wu[1], Jiangping Hu[1,4], Robert J Cava[2]†, Liling Sun[1,4]†

[1]Institute of Physics and Beijing National Laboratory for Condensed Matter Physics, Chinese Academy of Sciences, Beijing 100190, China
[2]Department of Chemistry, Princeton University, Princeton, New Jersey 08544, USA
[3]Institute for High Pressure Physics, Russian Academy of Sciences, 142190 Troitsk, Moscow, Russia
[4]University of Chinese Academy of Sciences, Beijing 100190, China
[5]Kavli Institute of Theoretical Sciences, University of Chinese Academy of Sciences, Beijing, 100049, China
[6]Institute of High Energy Physics, Chinese Academy of Sciences, Beijing 100049, China
[7]Shanghai Synchrotron Radiation Facilities, Shanghai Institute of Applied Physics, Chinese Academy of Sciences, Shanghai 201204, China
[8]National High Magnetic Field Laboratory, LANL, Los Alamos, New Mexico 87504, USA

We show that two different superconducting phases exist at high pressures in the optimized tetradymite topological insulators $Bi_2Te_2Se$ (BTS) and $Bi_{1.1}Sb_{0.9}Te_2S$ (BSTS). The superconducting phases emerge at structural phase transitions – the first at ~8.4 GPa for BTS and ~12.4 GPa for BSTS, and the second at 13.6 GPa for BTS and 20.4 GPa, for BSTS. Electronic structure calculations show that these phases do not have topological character. Comparison of our results with prior work on $Bi_2Se_3$, $Bi_2Te_3$ and $(Bi,Sb)_2(Se,Te)_3$ allows us to uncover a universal phase diagram for pressure-induced superconductivity in tetradymites, providing a basis for understanding the relationships between topological behavior, crystal structure, and superconductivity for these materials.

Key words: topological insulator, superconductivity, high pressure

**Introduction**

Tetradymites, well known as both thermoelectrics and topological insulators, have the general formula $M_2X_3$, in which M is a group V metal, usually Bi or Sb, and X is a group VI anion, Te, Se or S. These elements have similar electronegativites, and thus the materials favor band inversion [1]. Theoretical and experimental work has shown that the electrons on their surfaces are topologically protected and can display remarkable properties [2-15], thus making tetradymites one of the most important materials families for the study of topological insulators (TIs). If made superconducting, then they are of interest for applications ranging from spintronic to quantum computation [2,16-18] motivating us to investigate how they might be tuned from a topological insulating state to a superconducting state.

Chemical doping is a commonly available tuning method. It can produce disorder, defects and inhomogeneity in materials, however, and so is often not ideal. Pressure, on the other hand, is a clean way to realize the tuning of interactions in solids without introducing chemical complexity, and thus has been successfully adopted in the study of some TIs [19-33]. Our study of the evolution of the topological surface states on optimized tetradymite BTS and BSTS crystals (by optimized we mean highest bulk resistivities at ambient pressure) at high pressures has been reported recently, for example [34]. Here we focus on the investigation of the pressure-induced superconductivity in these two materials. We find that two pressure-induced superconducting phases appear after their bulk insulating states are suppressed. By comparing our results with available data for pressurized $Bi_2Se_3$ [21,33], $Bi_2Te_3$

[23,31,35] and (Bi,Sb)$_2$(Se,Te)$_3$ [28], we find a universal pressure dependent superconductivity phase diagram for all tetradymite TIs.

**Experimental details**

High-quality single crystals of BTS and BSTS were grown by the vertical Bridgman method, as described in Ref. [36,37]. Before the experiments, the crystals were freshly cleaved to expose pristine basal plane (001) surfaces.

The resistance measurements at high pressures were performed in a diamond anvil cell (DAC), in which diamond anvils with 400 $\mu$m flats and a nonmagnetic rhenium gasket with 100-$\mu$m-diameter hole were employed. The standard four-probe electrodes were applied on the cleavage plane of the BTS and the BSTS single crystals. To provide a quasi-hydrostatic pressure environment, NaCl powder was employed as the pressure medium. High-pressure X-ray diffraction (XRD) measurements were also performed in a DAC on beamline 4W2 at the Beijing Synchrotron Radiation Facility and on beamline 15U at the Shanghai Synchrotron Radiation Facility. Diamonds with low birefringence were selected for these XRD measurements. A monochromatic X-ray beam with a wavelength of 0.6199 Å was employed and silicon oil was employed as a pressure-transmitting medium. The pressure for all measurements in the DACs was determined by the ruby fluorescence method [38].

**Results and discussion**

In Fig.1 we show the electrical resistance as a function of temperature for BTS and BSTS at pressures up to 24.1 GPa and 31.8 GPa, respectively. As pressure increases, the resistances begin to drop near 2.1 K at about 8.4 GPa for BTS (Fig.1a and Fig.S1 in SI) and near 2 K at about 12.4 GPa for BSTS (Fig.1c and Fig.S1 in SI), The magnitude of these drops becomes more pronounced on further compression (Fig.1b and 1d), and is followed by sharp decreases to zero resistance at 10.7 GPa and 16 GPa (Fig.1d), respectively. The observation of zero resistance is a signature of superconductivity. The superconducting transition temperatures $T_C$ of these two materials exhibit identical pressure dependencies - increases upon elevating pressure (Fig.1b and 1d). Remarkably, the $T_C$s of BTS and BSTS display a sudden rise at 13.6 GPa and 20.4 GPa (see the red arrows in Fig.1b and 1d), which implies the appearance of a new superconducting phase [21,24]. On further increase of the pressure, the $T_C$ of the second superconducting phase decreases for both materials.

The superconductivity in these two TIs is confirmed by both high pressure measurements of the *ac* susceptibility and the resistance under magnetic fields; the former shows diamagnetic throws and the latter exhibits the magnetic field dependence of $T_C$ displayed by superconductors (Fig.S2 and Fig.S3 in the SI). In order to further differentiate the two superconductors, we estimate the upper critical magnetic field ($H_{c2}$) for the first and the second superconducting phases of BTS and BSTS by using the Werthamer-Helfand-Hohenberg (WHH) formula [39]: $H_{c2}^{WHH}(0) = -0.693 T_C (dH_{C2}/dT)_{T=Tc}$. Figure 2a and 2b present the plots of $H_{c2}$ versus $T_C$ obtained at different pressures for BTS and BSTS. The estimated values of the upper critical fields of BTS at zero temperature are ~ 0.34 T at 12.5 GPa and 1.96 T at 20.9 GPa (Fig.2a), while the $H_{c2}$ values of BSTS are 0.57 T at 12.6 GPa, ~0.94 T at 15.4 GPa and 2.27 T at 24.8 GPa (Fig.2b). The normalized critical field $h^*(t)$ (here $h^*(t) = [H_{C2}(T)/T_C]/[dH_{C2}(T)/dT]|_{T=T_C}$) as a function of $t = T/T_C$ is also plotted for the two materials, displayed in the inset of Fig.2. The data show that the $dh^*(t)/dt$ of the first superconducting phase of BTS subjected to 12.5 GPa has a different slope from that subjected to 20.9 GPa, consistently demonstrating that pressure indeed induces two distinct superconducting phases in BTS. Similar behavior is also observed in BSTS (inset of Fig.2b). The $dh^*(t)/dt$ of the sample measured at 15.4 GPa is different from that measured at 24.8 GPa, suggesting that the two TIs investigated in this study show the same kinds of changes under high pressure,

The electronic state of topological insulators is protected by time-reversal symmetry [1,3,40,41] and therefore structural stability is one of the key issues for understanding the superconductivity found in the pressure range of our experiments. It is known that BTS maintains its tetradymite structure to 8 GPa [29,42], but there are no reports of the high pressure structure of BSTS. Thus, we carried out high pressure X-ray diffraction measurements on BSTS. Figure 3a presents the X-ray diffraction patterns collected at pressures up to 36.1 GPa for BSTS. Like other tetradymite TIs, BSTS crystallizes in a rhombohedral (R) unit cell at ambient pressure [36,43] and maintains the R phase up to ~10.9 GPa. It then undergoes a structural phase transition at pressures between 10.9 and 13.1 GPa. The refinements for the high-pressure X-ray diffraction data collected at 17.3 GPa show that the high-pressure phase is monoclinic

(M) in space group C2/m (Fig.3b). On further increasing the pressure, BSTS converts into a tetragonal phase in space group I4/mmm at 22.7 GPa. The pressure dependence of lattice parameters in R, M and T phases are presented in Fig.3d.

To investigate whether the superconducting phases found in pressurized BTS and BSTS still possess a non-trivial topological nature, we calculated the band structures for the M and T phases for BTS based on the high pressure X-ray diffraction results [29]. Because the implied atomic disorder in the unit cell of BSTS makes the appropriate computations difficult, they were performed only for BTS. The calculations show that these two superconducting phases lose their topological nature due to the structural phase transitions under pressure (Fig.S4, Fig.S5 and Table 1 in the SI). The universal superconductivity behavior that we observe experimentally (see below) implies that the computed character of BTS can be applied for all tetradymite TIs, with differences in carrier concentration and disorder, of course.

We summarize our high pressure experimental results on BTS and BSTS in the pressure-temperature phase diagrams (Fig.4a and 4b). It is seen that these two TIs show the same type of behavior under pressure. There are three distinct ground states in the diagrams: the topological insulating state and the two superconducting states with distinct crystal structures. The first superconducting (SC-I) phase emerges in the monoclinic (M) phase, and its maximum $T_C$ is about 3.1 K for BTS and 6 K for the BSTS. At pressure above 17.1 GPa (BTS) and 22.1 GPa (BSTS), the second superconducting (SC-II) phase with tetragonal (T) structure is found. The maximum

$T_C$ value of the SC-II phase is almost two times higher than that of SC-I phase, indicating that the T phase has a higher $T_C$ in the these materials.

Finally, we compare the normalized $T_C$-*pressure* phase diagrams for $Bi_2Se_3$ $Bi_2Te_3$ and $(Bi,Sb)_2(Se,Te)_3$ with the results obtained in this study for BTS and BSTS. The normalization has been done using the observed phase transition pressures (The pressure of the rhombohedral to monoclinic transition for all materials is taken as 1 on the horizontal axis, and the maximum observed Tc for each material is taken as 1 on the vertical axis.) The phase diagrams reveal a remarkably universal character for all pressurized tetradymite TIs. As shown in Fig.S6, the SC-I phase for all the tetradymite TIs emerges at the appearance of the M phase and the SC-II phase develops at the appearance of the T phase. The critical pressures of the R to M and M to T transitions in different compounds are not the same, and the superconducting $T_C$s are not the same, but if we normalize the pressures to $P/P_{R-M}$, and the superconducting transition temperatures to $Tc/Tc_{max}$, then a clear universal superconductivity phase diagram is revealed for all the members of the tetradymite family. (Fig.4c).

**Conclusion**

In conclusion, we first find two pressure-induced superconducting phases in tetradymite topological insulators $Bi_2Te_2Se$ and $Bi_{1.08}Sn_{0.02}Sb_{0.9}Te_2S$ and demonstrate universal pressure dependent superconductivity phase diagrams for all the known tetradymite topological insulators. It is found that the SC-I phase emerges in a monoclinic phase and the SC-II phase appears in a tetragonal phase. The $T_C$ of the tetragonal superconducting phase is higher than that of the monoclinic superconducting phase. It is expected that the present work provides information that can lead to a unified understanding of the connection between crystal structure, topological character and superconductivity in pressurized tetradymites TIs.

**Acknowledgements**

We thank Profs. Hongming Weng, Qianghua Wang for helpful discussions. The work in China was supported by the National Key Research and Development Program of China (Grant No. 2017YFA0302900, 2016YFA0300300 and 2017YFA0303103), the NSF of China (Grants No. 11427805, No. U1532267, No. 11604376), the Strategic Priority Research Program (B) of the Chinese Academy of Sciences (Grant No. XDB07020300). The work at Princeton was supported by the ARO MURI on Topological Insulators, grant W911NF-12-1-0461.

Correspondence and requests for materials should be addressed to L. Sun (llsun@iphy.ac.cn) or R.J. Cava (rcava@Princeton.EDU)

**References**

[1]  J. P. Heremans, R. J. Cava, and N. Samarth, Nature Reviews Materials **2**, 17049 (2017).

[2]  M. Z. Hasan and C. L. Kane, Reviews of Modern Physics **82**, 3045 (2010).

[3]  Y. Xia *et al.*, Nature Physics **5**, 398 (2009).

[4] B. A. Bernevig, T. L. Hughes, and S.-C. Zhang, Science **314**, 1757 (2006).

[5] C. L. Kane and E. J. Mele, Physical Review Letters **95** (2005).

[6] M. König, S. Wiedmann, C. Brüne, A. Roth, H. Buhmann, L. W. Molenkamp, X.-L. Qi, and S.-C. Zhang, Science (2007).

[7] L. Fu, C. L. Kane, and E. J. Mele, Physical Review Letters **98** (2007).

[8] Y. Ando, Journal of the Physical Society of Japan **82**, 102001 (2013).

[9] X.-L. Qi and S.-C. Zhang, Reviews of Modern Physics **83**, 1057 (2011).

[10] Y. L. Chen *et al.*, Science **325**, 178 (2009).

[11] Y.-S. Fu, M. Kawamura, K. Igarashi, H. Takagi, T. Hanaguri, and T. Sasagawa, Nature Physics **10**, 815 (2014).

[12] D. Hsieh *et al.*, Nature **460**, 1101 (2009).

[13] D. Hsieh *et al.*, Physical Review Letters **103** (2009).

[14] K. Miyamoto *et al.*, Phys Rev Lett **109**, 166802 (2012).

[15] M. Neupane *et al.*, Physical Review B **85**, 235406 (2012).

[16] A. Y. Kitaev, Annals of Physics **303**, 2 (2003).

[17] F. Wilczek, Nature Physics **5**, 614 (2009).

[18] C. Nayak, S. H. Simon, A. Stern, M. Freedman, and S. Das Sarma, Reviews of Modern Physics **80**, 1083 (2008).

[19] X. Xi, C. Ma, Z. Liu, Z. Chen, W. Ku, H. Berger, C. Martin, D. B. Tanner, and G. L. Carr, Physical Review Letters **111** (2013).

[20] Y. Zhou *et al.*, Physical Review B **93** (2016).

[21] K. Kirshenbaum *et al.*, Physical Review Letters **111** (2013).

[22] R. Vilaplana *et al.*, Physical Review B **84** (2011).

[23] L. Zhu, H. Wang, Y. Wang, J. Lv, Y. Ma, Q. Cui, Y. Ma, and G. Zou, Physical Review Letters **106** (2011).

[24] R. L. Stillwell, Z. Jenei, S. T. Weir, Y. K. Vohra, and J. R. Jeffries, Physical Review B **93** (2016).

[25] T. V. Bay, T. Naka, Y. K. Huang, H. Luigjes, M. S. Golden, and A. de Visser, Physical Review Letters **108** (2012).

[26] G. Liu, L. Zhu, Y. Ma, C. Lin, J. Liu, and Y. Ma, The Journal of Physical Chemistry C **117**, 10045 (2013).

[27] Y. Ma, G. Liu, P. Zhu, H. Wang, X. Wang, Q. Cui, J. Liu, and Y. Ma, Journal of Physics: Condensed Matter **24**, 475403 (2012).

[28] J. R. Jeffries, N. P. Butch, Y. K. Vohra, and S. T. Weir, Journal of Physics: Conference Series **592**, 012124 (2015).

[29] J. Zhao *et al.*, Physical Chemistry Chemical Physics **19**, 2207 (2017).

[30] J. R. Jeffries, A. L. Lima Sharma, P. A. Sharma, C. D. Spataru, S. K. McCall, J. D. Sugar, S. T. Weir, and Y. K. Vohra, Physical Review B **84** (2011).

[31] K. Matsubayashi, T. Terai, J. S. Zhou, and Y. Uwatoko, Physical Review B **90** (2014).

[32] A. M. Nikitin, Y. Pan, Y. K. Huang, T. Naka, and A. de Visser, Physical Review B **94** (2016).

[33] Z. Yu, L. Wang, Q. Hu, J. Zhao, S. Yan, K. Yang, S. Sinogeikin, G. Gu, and H.-k. Mao, Scientific Reports **5** (2015).

[34] Shu Cai, Jing Guo. Vladimir A. Sidorov, Yazhou Zhou, Honghong Wang, Gongchang Lin, Xiaodong Li, Yanchuan Li, Ke Yang, Aiguo Li, Qi Wu, Jiangping Hu, S. K. Kushwaha, Robert J Cava, Liling Sun ArXiv e-prints (2018).

[35] C. Zhang, L. Sun, Z. Chen, X. Zhou, Q. Wu, W. Yi, J. Guo, X. Dong, and Z. Zhao, Physical Review B **83** (2011).

[36] S. K. Kushwaha *et al.*, Nature Communications **7**, 11456 (2016).

[37] S. Jia, H. Ji, E. Climent-Pascual, M. K. Fuccillo, M. E. Charles, J. Xiong, N. P. Ong, and R. J. Cava, Physical Review B 84 (2011).

[38] H. K. Mao, J. Xu, and P. M. Bell, Journal of Geophysical Research **91**, 4673 (1986).

[39] N. R. Werthamer, E. Helfand, and P. C. Hohenberg, Physical Review **147**, 295 (1966).

[40] H. Zhang, C.-X. Liu, X.-L. Qi, X. Dai, Z. Fang, and S.-C. Zhang, Nature Physics **5**, 438 (2009).

[41] L. Fu and C. L. Kane, Physical Review B **76** (2007).

[42] M. B. Nielsen, P. Parisiades, S. R. Madsen, and M. Bremholm, Dalton Trans **44**, 14077 (2015).

[43] T. Misawa, Y. Fukuyama, Y. Okazaki, S. Nakamura, N. Nasaka, T. Sasagawa, and N. H. Kaneko, IEEE Transactions on Instrumentation and Measurement **66**, 1489 (2017).

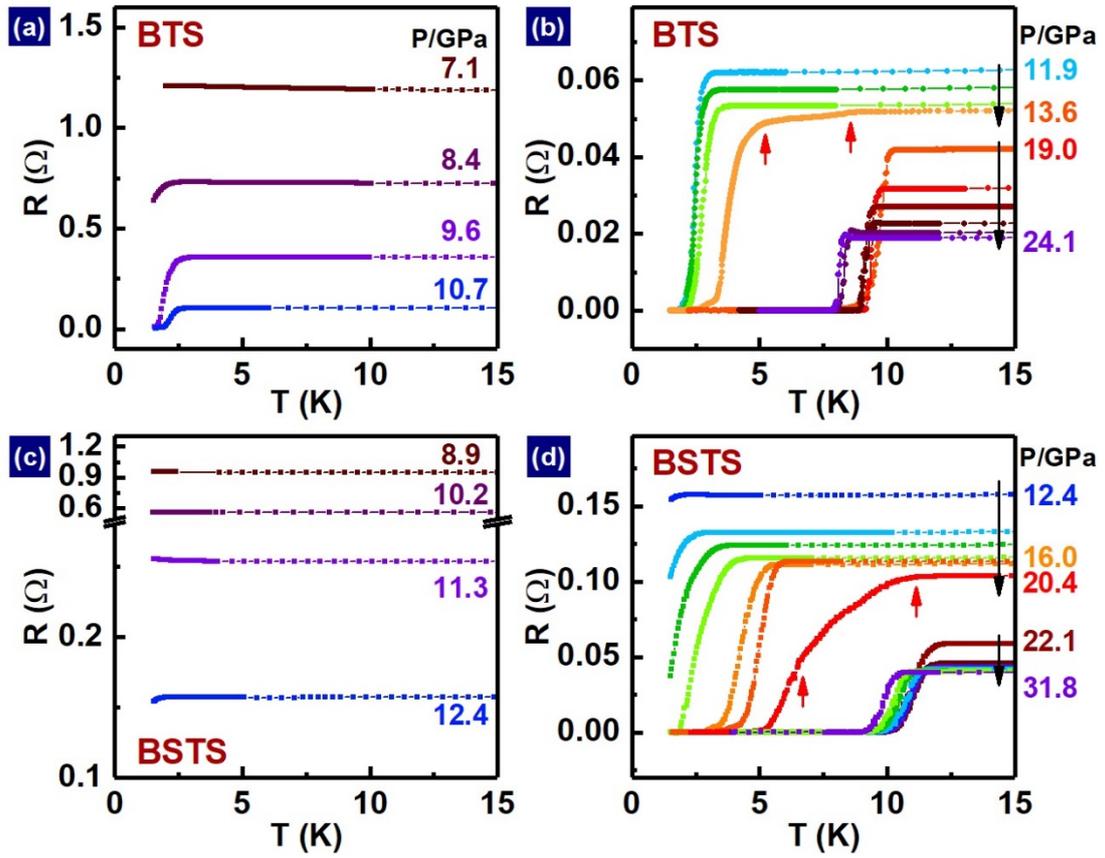

**Figure 1 Transport properties of BTS and BSTS at different pressures below 15 K.** (a) and (b) display temperature dependence of electrical resistance obtained at different pressures for BTS. (c) and (d) show resistance as a function of temperature measured at different pressures for BSTS.

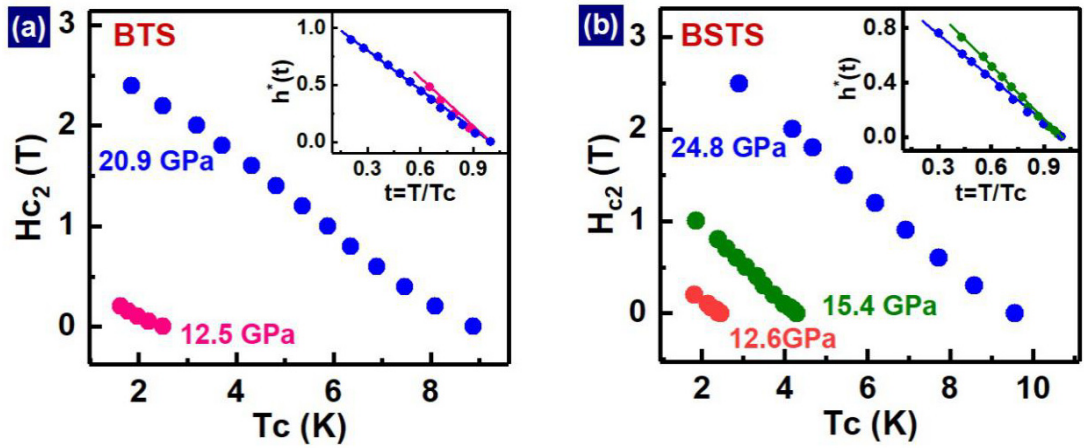

**Figure 2 Upper critical field $H_{C2}$ as a function of superconducting transition temperature $T_C$ for pressurized BTS and BSTS.** The insets display corresponding normalized critical field $h^*$ as a function of $t = T/T_C$.

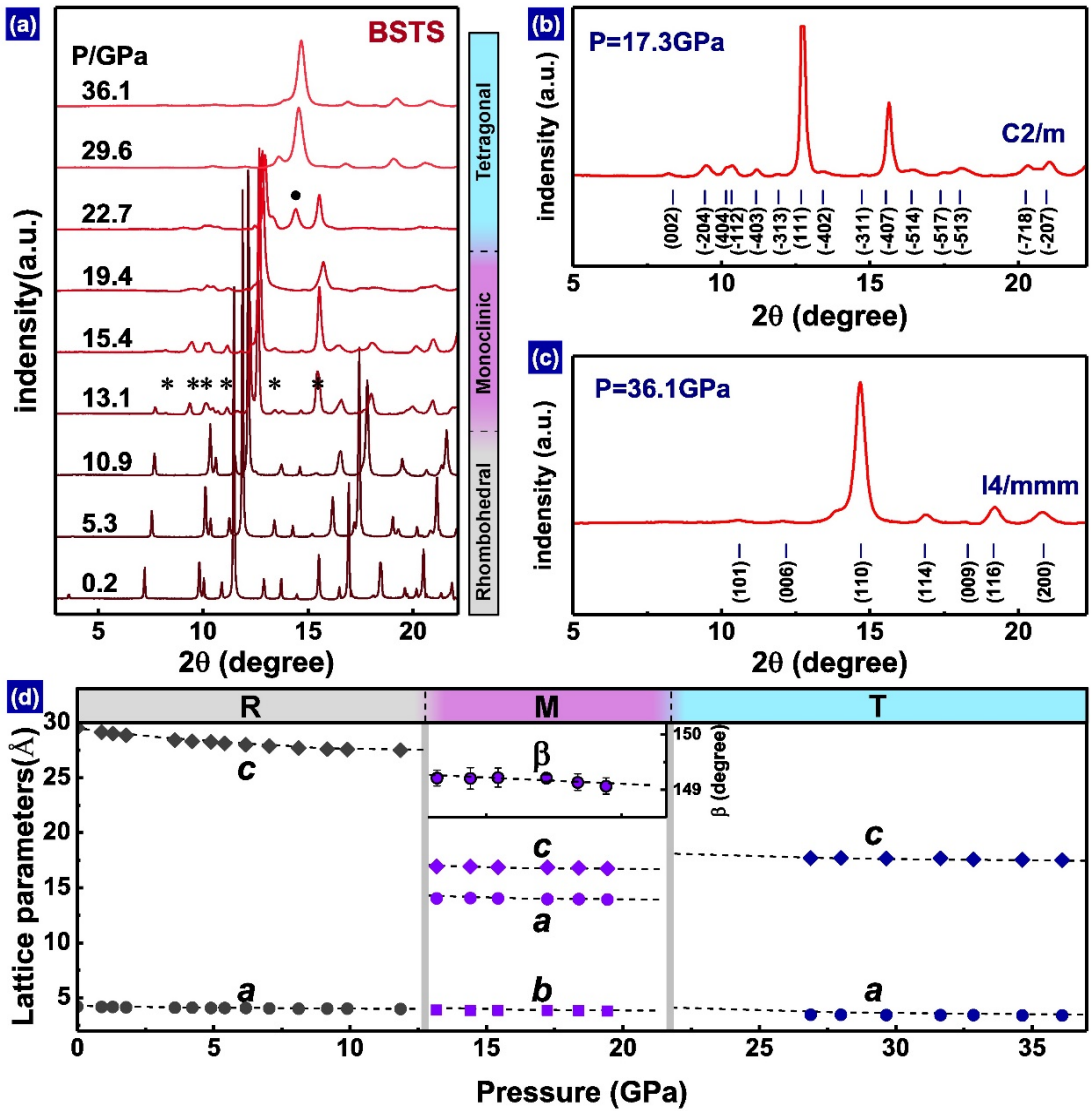

**Figure 3 Structural information for pressurized BSTS.** (a) X-ray diffraction patterns collected at different pressures, displaying pressure-induced phase transitions at 13.1 GPa and 22.7 GPa. Stars (*) and black solid in the figure indicate new peaks. (b) and (c) Refinements results on the new phases at 17.3 GPa and 36.1GP, illustrating that they crystallize in monoclinic and tetragonal unit cells, respectively. (d) Pressure dependence of lattice parameters for rhombohedral (R), monoclinic (M) and tetragonal (T) phases.

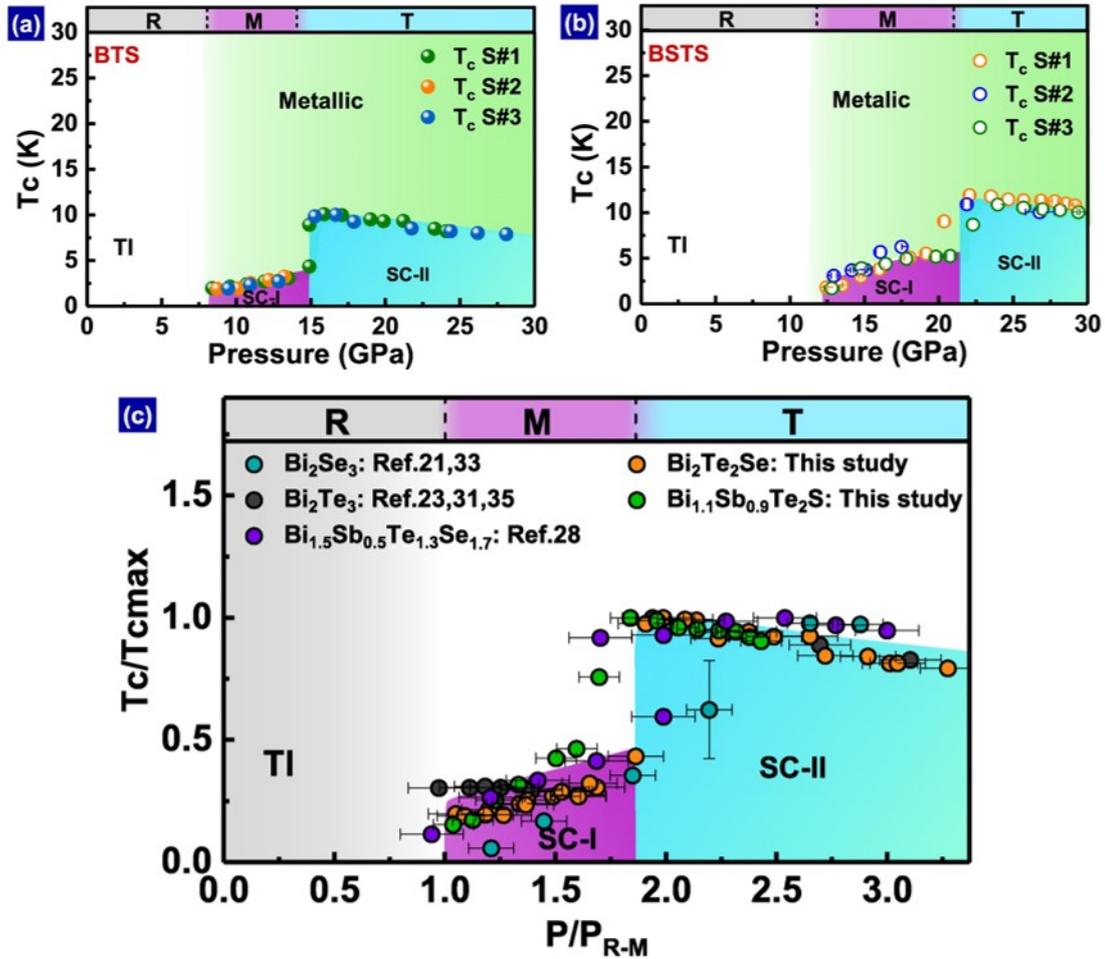

**Figure 4 Pressure-temperature phase diagrams of tetradymite TIs.** (a) and (b) Pressure versus superconducting transition temperature $T_C$ for BTS and BSTS, respectively. The acronym TI stands for the topological insulating state. SC-I and SC-II represent superconducting states with distinct crystal structures. R, M and T stand for rhombohedral, monoclinic and tetragonal phases, respectively. S#1, S#2 and S#3 represent samples 1, sample 2 and sample 3. (c) Plot of $P/P_{R-M}$ and $T_C/T_{Cmax}$ for $Bi_2Te_3$, $Bi_2Se_3$, $(Bi,Sb)_2(Se,Te)_3$, BTS and BSTS, displaying a universal behavior for the tetradymite TIs.

# Supplementary Information for "Universal pressure dependent superconductivity phase diagrams for tetradymite topological insulators"

Shu Cai[1,4], S. K. Kushwaha[2,8], Jing Guo[1], Vladimir A. Sidorov[3], Congcong Le[1,5], Yazhou Zhou[1], Honghong Wang[1,4], Gongchang Lin[1,4], Xiaodong Li[6], Yanchuan Li[6], Ke Yang[7], Aiguo Li[7], Qi Wu[1], Jiangping Hu[1,4], Robert J Cava[2]†, Liling Sun[1,4]†

[1]*Institute of Physics and Beijing National Laboratory for Condensed Matter Physics, Chinese Academy of Sciences, Beijing 100190, China*

[2]*Department of Chemistry, Princeton University, Princeton, New Jersey 08544, USA*

[3]*Institute for High Pressure Physics, Russian Academy of Sciences, 142190 Troitsk, Moscow, Russia*

[4]*University of Chinese Academy of Sciences, Beijing 100190, China*

[5]*Kavli Institute of Theoretical Sciences, University of Chinese Academy of Sciences, Beijing, 100049, China*

[6]*Institute of High Energy Physics, Chinese Academy of Sciences, Beijing 100049, China*

[7]*Shanghai Synchrotron Radiation Facilities, Shanghai Institute of Applied Physics, Chinese Academy of Sciences, Shanghai 201204, China*

[8]*National High Magnetic Field Laboratory, LANL, Los Alamos, New Mexico 87504, USA*

†Correspondence and requests for materials should be addressed to L. Sun (llsun@iphy.ac.cn) or R.J. Cava (rcava@Princeton.EDU)

1. **High-pressure resistance and *ac* susceptibility measurements on our BTS and BSTS**

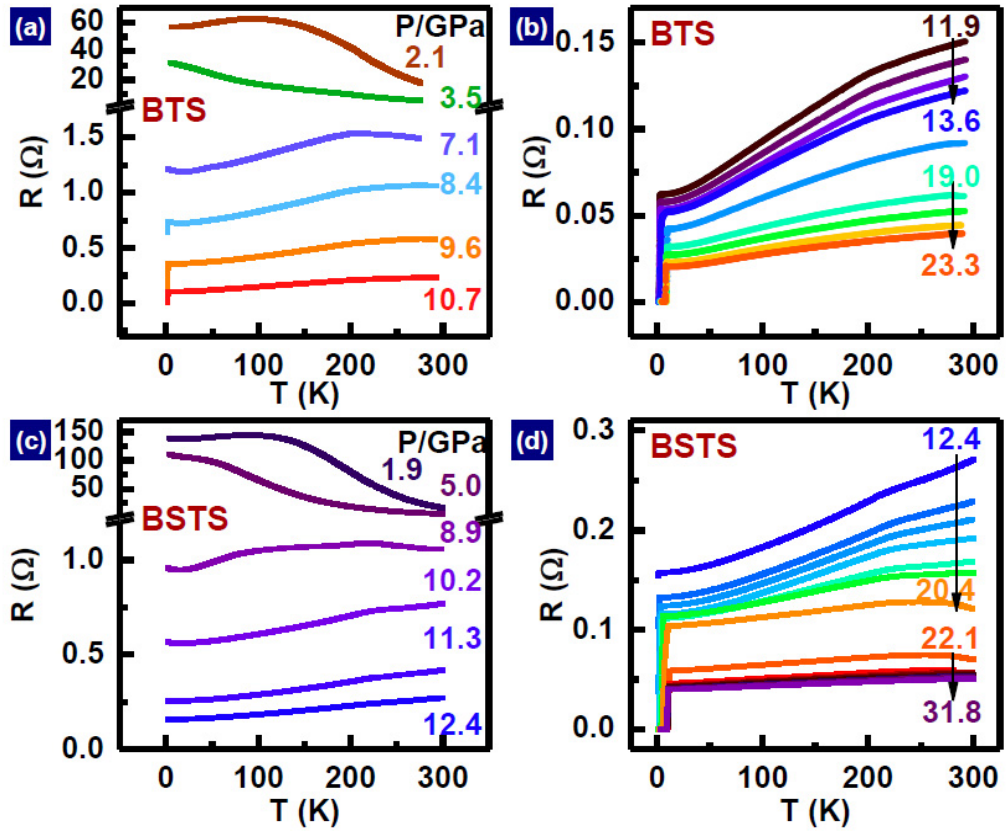

**Figure S1** (a) and (b) Temperature dependence of resistance in BTS for pressures ranging from 2.1 GPa to 23.3 GPa. (c) and (d) Resistance as a function of temperature in BSTS for pressures ranging from 1.9 GPa to 31.8 GPa.

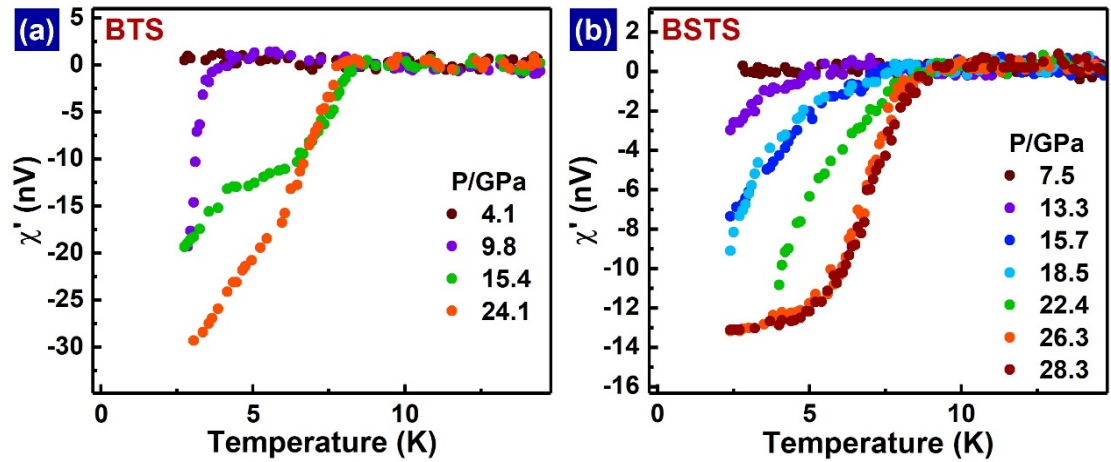

**Figure S2** Results of high-pressure *ac* susceptibility measurements on BTS and BSTS, displaying diamagnetic throws at pressures where the samples show the zero resistance.

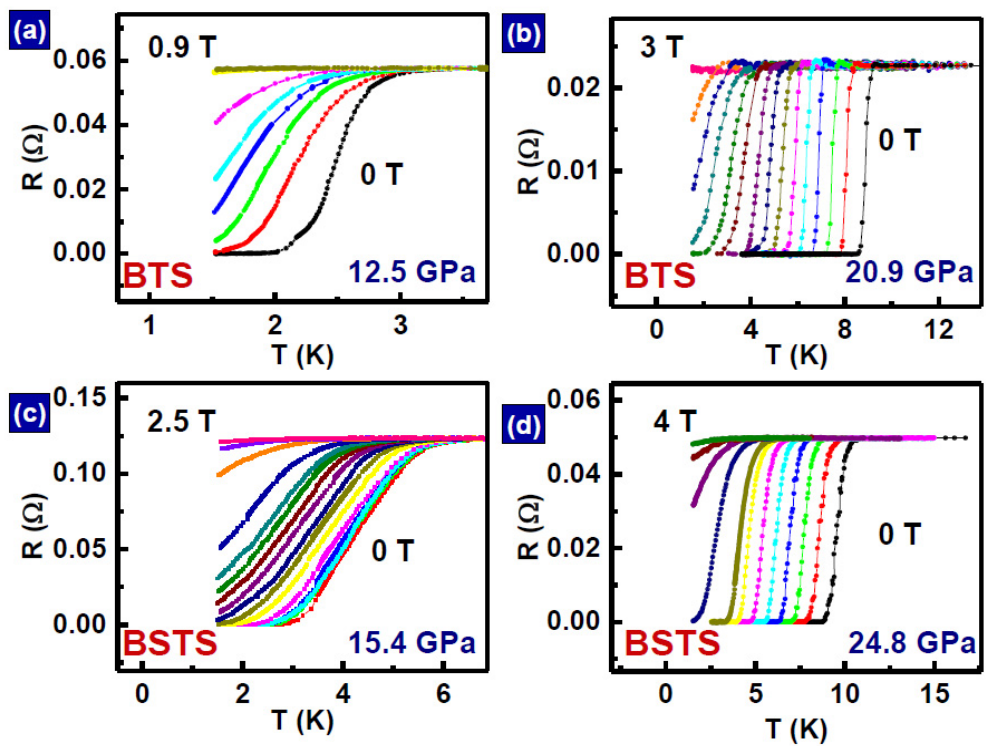

**Figure S3** Magnetic field dependence of superconducting transition temperatures in BTS (see figure a and figure b) and BSTS (see figure c and figure d).

## 2 The determination of high pressure crystal structure for BSTS

As can be seen in Fig3, BSTS crystallizes in a rhombohedral (R) unit cell at ambient pressure and undergoes a structural phase transition at pressures between 10.9 and 13.1 GPa. The peaks can be indexed to a monoclinic phase with crystal parameters similar to other tetradymite family materials, for instance, $Bi_2Te_3$, $Bi_2Se_3$ and $Sb_2Te_3$[1-9]. This monoclinic phase remains the crystal structure below 21 GPa.

At ~21 GPa, new diffraction peaks appear, indicating that the sample undergoes a phase transition. The pressure-induced phase in BSTS was observed in other tetradymite materials [1-3,5-9], in which this phase is always seen in $Bi_2Te_3$ and $Sb_2Te_3$ and is indexed as a body-centered cubic (BCC) phase [2,6,9]. However, we note that the higher pressure phase of $Bi_2Se_3$ and BSTS can be indexed better as a tetragonal phase in space group I4/mmm than the BCC phase [1,5,7], and find that the tetragonal phase of BSTS can be described by five BCC-like unit cells arranged along the *c* axis (Fig.S5a). The large difference of the atom radius in BSTS and $Bi_2Se_3$ seems to be in favor of the formation of the tetragonal structure which is composed of BCC-like sub-units.

## 3 Theoretical calculations on the band structures of the monoclinic and the tetragonal phases

### 3.1 Methods

Our density functional theory (DFT) calculations are performed with the Vienna ab initio simulation package (VASP) code [10-12]. The Perdew-Burke-Ernzerhof (PBE) exchange-correlation functional and the projector augmented-wave (PAW) approach are used. Throughout the work, the cutoff energy is set to be 450 eV for expanding the wave functions into plane-wave basis. In the calculation, the Brillouin zone is sampled in the k space within Monkhorst Pack scheme [13]. On the basis of the equilibrium structure, the k mesh used is $6 \times 6 \times 5$ and $5 \times 5 \times 5$ for primitive monoclinic and tetragonal unit cell, respectively.

### 3.2 Crystal and band structures

In the main text, the crystal chemistry of $Bi_2Te_2Se$ is characterized by three stable structural phases (rhombohedral, monoclinic and tetragonal phase) under the different pressure. Figure S4a and Figure S5a show the experimental crystal structures of the monoclinic and tetragonal phases. The lattice parameters and atom positions for the monoclinic and tetragonal phase are described in Ref.8. In the rhombohedral phase, $Bi_2Te_2Se$ is a topological insulator [14,15]. In order to investigate the topological properties of the monoclinic and tetragonal phases, the band structures with spin orbital coupling (SOC) are calculated, shown in Fig.S4b and Fig.S5b. The two phases are metallic, which consists with our experimental results. In the monoclinic phase, near the Fermi level the bands are mainly attributed to the Bi-6p and Te-5p orbitals and the gap exists between the valence and conduction bands, shown in Fig.S4b. To further confirm the topological properties of the monoclinic $Bi_2Te_2Se$, we adopt the parity check method proposed by Fu and Kane [16]. Parities at all time-reversal invariant momenta in SOC bands of monoclinic $Bi_2Te_2Se$ are listed in Table.S1, which indicates that the band structure is topological trivial, implying that the sample lost its non-trivially topological state in the monoclinic superconducting phase. It is important to note that the primitive cell contains two formula units so that Te-5p and Se-4p orbitals can contribute to 18 double degenerate bands. We also compute the tetragonal phase and find that there is no gap between the valence and conduction bands (Fig.S5), suggesting that the tetragonal superconducting phase also shows no non-trivially topological character.

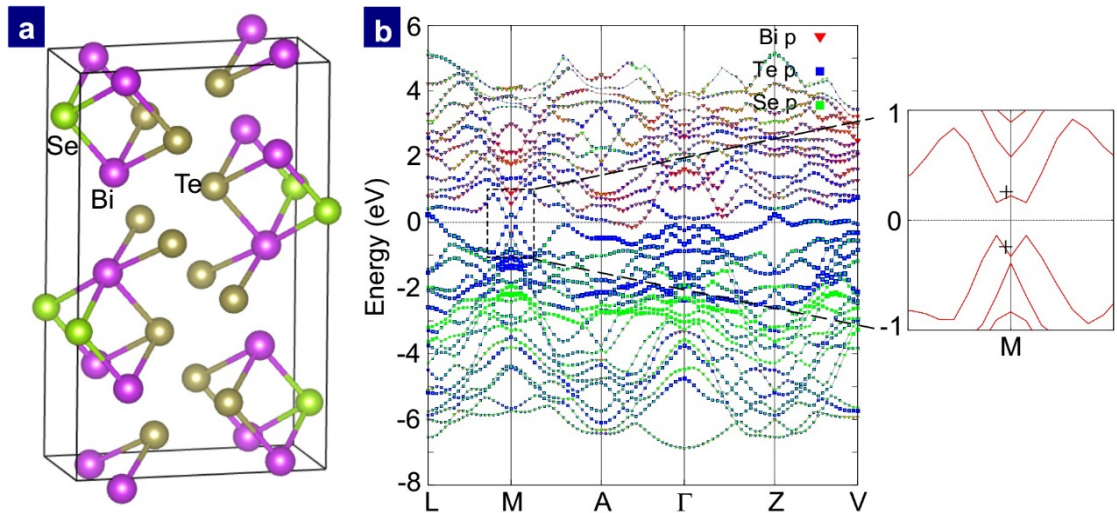

**Figure S4** (a) Crystal structure of the monoclinic phase of BTS sample. (b) Corresponding band structure.

TABLE S1: Parities for each band at all Time-reversal invariant momenta(TRIM). There are 18 double degenerate bands.

| Parities(SOC) \ band No TRIM | 1 | 2 | 3 | 4 | 5 | 6 | 7 | 8 | 9 | 10 | 11 | 12 | 13 | 14 | 15 | 16 | 17 | 18 |
|---|---|---|---|---|---|---|---|---|---|---|---|---|---|---|---|---|---|---|
| (0 0 0) | + | - | + | + | - | - | + | + | - | + | - | + | - | + | - | + | - | - |
| (0.5 0.5 0.5) | + | - | - | + | + | - | + | - | + | - | - | - | + | + | - | + | - | + |
| (0.5 0.5 0) | - | + | - | + | + | - | + | - | + | - | - | + | - | + | + | + | - | - |
| (0.5 0 0.5) | + | - | - | + | - | + | - | - | + | + | - | + | + | - | + | + | - | - |
| (0 0.5 0.5) | - | + | - | + | + | - | + | - | + | - | - | + | - | + | + | + | - | - |
| (0 0 0.5) | - | + | - | + | + | - | + | - | + | - | - | + | + | - | + | - | + | - |
| (0 0.5 0) | - | + | - | + | - | + | + | - | + | + | - | - | + | + | - | + | - | - |
| (0.5 0 0) | - | + | - | + | + | - | + | - | + | - | - | + | + | - | + | - | + | - |
| $Z_2$(TI) | | | | | | | | | | | | | | | (0;000) | | | |

**Table S1** Parities for each band at all time-reversal invariant momenta (TRIM).

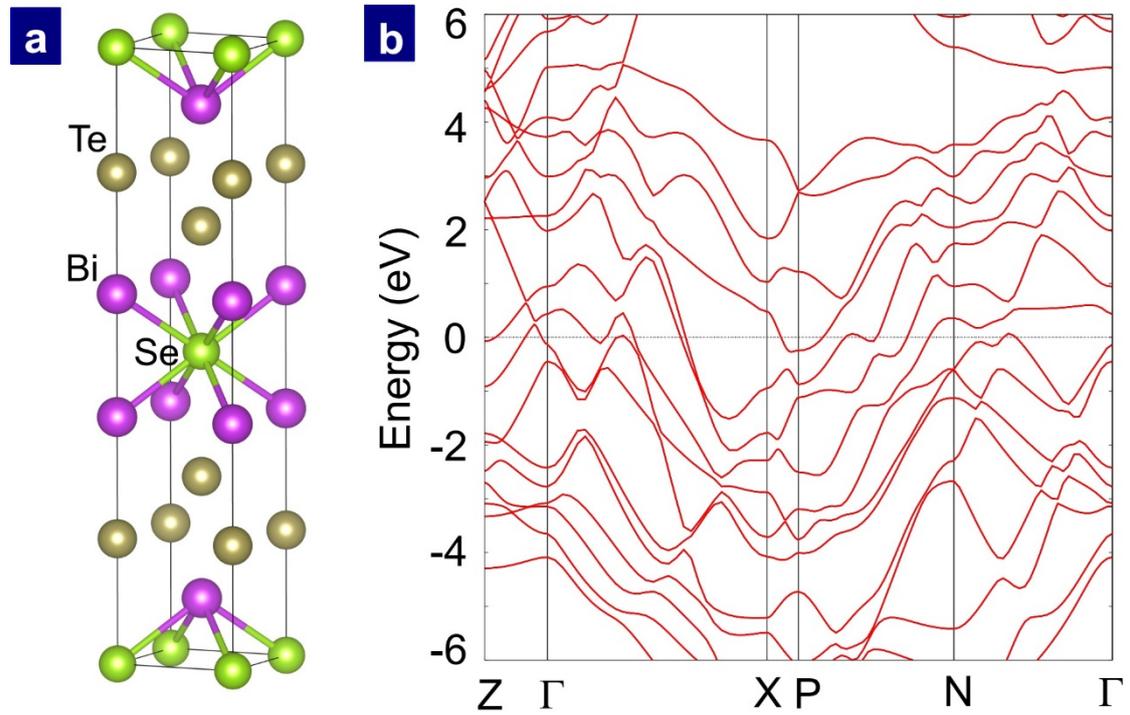

**Figure S5** (a) Crystal structure of the tetragonal phase of BTS. (b) Corresponding band structure.

**4 Universal pressure dependent superconductivity in tetradymite topological insulators**

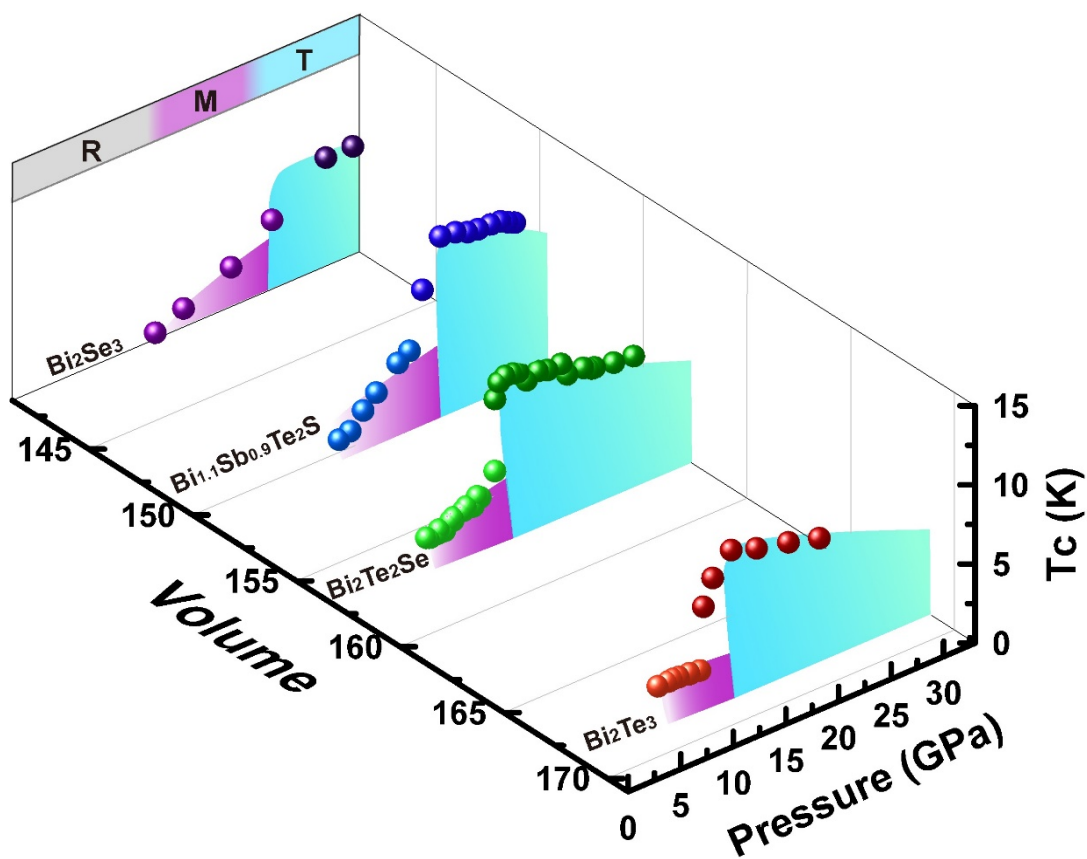

**Figure S6** Pressure/volume dependence of superconductivity phase diagram for the tetradymite topological insulators, showing a universal behavior under pressure.

**References**

[1]  G. Liu, L. Zhu, Y. Ma, C. Lin, J. Liu, and Y. Ma, The Journal of Physical Chemistry C **117**, 10045 (2013).

[2]  Y. Ma, G. Liu, P. Zhu, H. Wang, X. Wang, Q. Cui, J. Liu, and Y. Ma, Journal of Physics: Condensed Matter **24**, 475403 (2012).

[3]  M. B. Nielsen, P. Parisiades, S. R. Madsen, and M. Bremholm, Dalton Trans **44**, 14077 (2015).

[4]  R. Vilaplana *et al.*, Physical Review B **84** (2011).

[5]  Z. Yu, L. Wang, Q. Hu, J. Zhao, S. Yan, K. Yang, S. Sinogeikin, G. Gu, and H.-k. Mao, Scientific Reports **5** (2015).

[6]  J. Zhao, H. Liu, L. Ehm, Z. Chen, S. Sinogeikin, Y. Zhao, and G. Gu, Inorganic Chemistry **50**, 11291 (2011).

[7]  J. Zhao, H. Liu, L. Ehm, D. Dong, Z. Chen, and G. Gu, Journal of Physics: Condensed Matter **25**, 125602 (2013).

[8]  J. Zhao *et al.*, Phys. Chem. Chem. Phys. **19**, 2207 (2017).

[9]  L. Zhu, H. Wang, Y. Wang, J. Lv, Y. Ma, Q. Cui, Y. Ma, and G. Zou, Physical Review Letters **106** (2011).

[10] G. Kresse and J. Hafner, Physical Review B **47**, 558 (1993).

[11] G. Kresse and J. Furthmüller, Physical Review B **54**, 11169 (1996).

[12] G. Kresse and J. Furthmüller, Computational Materials Science **6**, 15 (1996).

[13] H. J. Monkhorst and J. D. Pack, Physical Review B **13**, 5188 (1976).

[14] S. K. Kushwaha, Q. D. Gibson, J. Xiong, I. Pletikosic, A. P. Weber, A. V. Fedorov, N. P. Ong, T. Valla, and R. J. Cava, Journal of Applied Physics **115**, 143708 (2014).

[15] J.-L. Mi *et al.*, Advanced Materials **25**, 889 (2013).

[16] L. Fu and C. L. Kane, Physical Review B **76** (2007).